\documentclass[12pt]{article}
\usepackage{amsmath}
\usepackage{amssymb}
\usepackage{graphicx}
\usepackage[cp1251]{inputenc}
\newcommand{\be}{\begin{equation}}
\newcommand{\ee}{\end{equation}}
\newcommand{\ba}{\begin{eqnarray}}
\newcommand{\ea}{\end{eqnarray}}
\newcommand{\ds}{\displaystyle}
\newcommand{\fs}{\footnotesize}

\begin{document}

\vspace{1cm}

\begin{center}
 \bf Conductance of Finite-Scale Systems with Multiple Percolation Channels
\end{center}
\bigskip

\centerline{E.Z. Meilikhov$^*$}
\medskip
\centerline{\small\it RRC “Kurchatov Institute”, 123182 Moscow,
Russia}
  \vspace{1cm}

  \begin{center}
\begin{tabular}{p{15cm}}
\footnotesize \hspace{15pt} We investigate properties of
two-dimensional finite-scale percolation systems whose size along
the current flow is smaller than the perpendicular size. Successive
thresholds of appearing multiple percolation channels in such
systems have been determined and dependencies of the conductance on
their size and percolation parameter $p$ have been calculated.
Various experimental examples show that the finite-scale percolation
system is the natural mathematical model suitable for the
qualitative and quantitative description of different physical
systems. \vspace{5mm}

\end{tabular}
\end{center}

\centerline{\bf 1. Introduction}
 \medskip

The percolation threshold $p_{c\infty}$ for the {\it infinite}
percolation system is the quite definite value depending on the
system topology and dimension. When the percolation parameter $p$
defining the fraction of conducting regions becomes higher, the
\emph{infinite} conducting cluster arises whose conductance is the
smooth function of $p$~\cite{1}. Recently, the new remarkable
results have been obtained for infinite rectangular lattices at
critical value of $p$. Firstly, it has been found numerically that
the probability of appearing the spanning cluster is an universal
function the rectangular ratio only~\cite{2}. Secondly, it has been
proved that more than one spanning cluster could exist in
two-dimensional critical percolation systems~\cite{3}, and, third,
it has been shown by numerical calculations that there could be more
than one cluster spanning the short direction of a long
strip~\cite{4}. The exact formulae (the crossing formulae) for the
probability that a certain number of crossing clusters (channels)
span a long strip has been derived in~\cite{5}.

However, all cited results refer to \emph{infinite} systems only (or
to the finite systems with lattice cell approaching zero which are
equivalent to the infinite system). For the {\it finite-scale}
system, the situation changes: the percolation threshold $p_c$
corresponds to the formation of the {\it finite-scale} crossing
cluster that connects opposite sides of the system. The threshold
$p_c$ becomes to be random quantity taking different values for
various realizations of the system~\cite{6,7}. In that case, one
could say about the probability to get one or another value  of
$p_c$ only or about the distribution function $f(p_c)$ of
percolation thresholds for an ensemble of systems of the certain
topology and geometry. For the infinite system
$f(p_c)=\delta(p_c-p_{c\infty})$; for the finite-scale system, the
distribution function becomes broader and the average threshold
value $\bar p_c=\int p_c f(p_c)dp_c$ shifts (in one or another
direction) relative to $p_{c\infty}$.

Finite-scale percolation system is the natural mathematical model of
various physical systems: 1) two-dimensional electron gas near the
surface of semiconductor with spatial potential fluctua\-tions due
to a non-uniform distribution of impurities~\cite{8}, 2) cluster
metal film whose resistance changes in the course of the
deposition~\cite{9,10}, 3) hydrogenated amorphous silicon with its
unusual noise features~\cite{11,12} and even  lightning
discharges~\cite{13} (see below). In the most cases, the $p$-value
is determined by a certain physical parameter that could be
controlled. For instance, in non-uniform semiconductor the $p$-value
is determined by Fermi level whose position could be changed by
means of the gate potential, in cluster film the $p$-value varies
temporally in the course of the film growth, in \emph{a}-Si it is
defined by the hydrogen content. Therefore, various features of
above-mentioned physical phenomena could be described by
investigating percolation peculiarities for the respective model
systems. \vspace{0.5cm}

\centerline{\bf 2. Percolation in finite-scale systems}
 \medskip

For the system of square form, the distribution function
$f_\Box(p_c)$ of thresholds is close to the Gaussian
one\footnote{More accurate calculations show the distribution
function is slightly asymmetric and is not Gaussian (see Fig. 1).
Moreover, its distinction from Gaussian function does not disappear
even with increasing the lattice size ~\cite{14}.}~\cite{15}. For
the rectangular system characterized by the aspect ratio $L/h$ ($h$
is the system “height”, $L$ is its “length”) the distribution
function $f(p_c)$ remains to be near-Gaussian, however the position
of its maximum and the width are significantly changed. If the
distribution function $f_\Box(p_c)$ of thresholds for the square
($L/h=1$) is known, then the distribution function for the system
with $L\gg h$ could be found by means of the approximate relation
 \be\label{0}
f(p_c)\approx(L/h)f_\Box(p_c)\left[1-\int_0\limits^{p_c}
f_\Box(p_c)dp_c\right]^{L/h-1}.
 \ee

Keeping in mind to study the  conductance of percolation lattices
similar to the long strip ($L\gg h$) and to use parameters of the
relevant functions $f(p_c)$ for the description of conductance
features (mostly, for the current flowing along $h$), we have
performed numerical investigating the site percolation problem for
the system comprising the finite-scale rectangular lattice with
square cells. The ratio of the lattice height $h$ and length $L$
(being measured by the numbers of cells contained, respectively,
along and transverse to the mean direction of the current flow
through the lattice) has been varied in the broad enough range to
study the percolation not only in the lattice of the square form but
also along ($L/h\ll 1$) and across ($L/h\gg 1$) the narrow  $h\times
L$-“strip”. Early, similar calculations (but for square systems of
rather large sizes, $h=L>32$, only and in the immediate vicinity of
$p_c$) have been performed in~\cite{16,17}. The percolation
probability for rectangular systems of different aspect ratios for
fixed $p$-values has been calculated in~\cite{18}.

Results of percolation calculations for the site problem are
presented in Figs. 1, 2 and lead to the following conclusions:
\begin{itemize}
\item the mean percolation threshold $\bar p_c$  for the {\it square}
lattice is close to the percolation threshold $p_{c\infty}$ of the
infinite lattice (in our case, $p_{c\infty}\approx 0.59$);

\item the mean percolation threshold {\it across} the narrow
 “strip” is significantly  {\it lower} than $p_{c\infty}$;

\item the mean percolation threshold {\it along} the narrow “strip” is
{\it higher} than $p_{c\infty}$, with $\bar p_c\to 1$ at $L/h\to
0$;

\item distribution functions of percolation thresholds are close
to the Gaussian function  $f(p_c)\propto\exp[(p_c-\bar
p_c)^2/2\sigma^2]$;

\item the width $\sigma$ of the distribution function $f(p_c)$ (defined
by double standard deviation  $2\sigma$) is maximum for the
square lattice;

\item approximate formulae (\ref{0}) leads to the  correct
(with accuracy better than 10\%)  parameter values of approximating
Gaussian function\footnote{It is the consequence of the known fact
that transverse and longitudinal sizes of the cluster arising at the
percolation threshold are nearly the same (about of
$h$)~\cite{19})}.

\end{itemize}

In addition to the significant spread of percolations thresholds,
another feature of the finite-scale percolation system is the
possibility of the simultaneous existence of several clusters
connecting its opposite sides and representing separate percolation
paths~\cite{3,20}. Here, the system conductance $G(p)$ ceases to be
the smooth function of $p$ and increases (with $p$) by more or less
pronounced jumps corresponding to appearing new percolation paths.
It is illustrated by Fig. 3 where the calculated dependence $G(p)$
for the transverse conductance of the long $10\times1000$-strip on
the $p$-value is shown (naturally, that is the result relative to
one of numerous random realizations of such a system). The relevant
numerical calculations for the site problem have been performed with
applying the known transfer matrix method~\cite{21} for the lattice
where the conductance of bonds between occupied sites (which
fraction equals $p$) has been assumed to be $G_0=1$ Ohm$^{-1}$, and
conductances of all other bonds -- to be $G_{\rm min}=10^{-5}G_0$.
Each jump of the conductance is on the order of value equal to 0.1
Ohm$^{-1}=G_0/10$. In consideration of $h=10$, it means that
successively appeared new conducting channels are isolated from each
other and almost linear.

Considering the dependence $G(L)$ at a given $p$-value one could
recognize how many percola\-tion paths (parallel conducting
channels) exist  in the long strip $L\times h$ and how far are they
from each other at that $p$-value. Indeed, the every sharp increase
of the conductance $G(L)$ is the sign of appearing the new
percolation channel. Let such jumps happen at the values $L_i$
($i$=1, 2,…). Then the difference  $L_{i+1} -L_i$  is the distance
between $(i+1)$-th  and $i$-th channels. It is just this dependence
$G(L)$ which is determined naturally in the course of the
conductance calculations by transfer matrix method. Several such
dependencies (corresponding to the same random realization of
$10\times1000$-system as for Fig. 2) are displayed in Fig. 4. They
show that the first conducting channel arises at $p\approx0.3$, and
already at $p\approx0.45$ the number of such channels come up to
ten, with their distribution over the system length being more or
less uniform. However, even at $p\approx0.5< p_{c\infty}$ those
channels merge and the current
 through the strip flows uniformly.

Percolation paths originating with increasing $p$ could be of two
kinds: they are either connected with the “old” paths (having
aroused at lower values of the parameter $p$) or isolated completely
from those, and then one could say about arising new percolation
channel. The highest conductance jumps could be naturally associated
with the percolation paths of the second type. Therewith, one could
say about the successive (as $p$ becomes higher) “switching”
\textit{different} percolation channels in the system. Obviously,
that scenario is mostly probable in the systems whose width $h$
(measured along the direction of the current flow) is much less than
the length $L$. In what follows, we shall consider just that system
and search the most probable values of the successive “percolation
thresholds” $p_c^{(n)}$ corresponding to originating new percolation
channels.

If the distribution function $f(p_c)$ for percolation thresholds
(with the standard meaning of this term relating to originating the
first percolation path) is known for a given system (for instance,
by numerical modeling, see Fig. 1), then the sequence $\bar
p_c^{(n)}$ ($n=1,2,\ldots$) of the mean values of percolation
thresholds could be found in the following way. The probability of
falling the  next ($n$-th) threshold within the interval from
$p_c^{(n)}$ to $p_c^{(n)}+dp_c$ equals
 \be\label{1}
f[p_c^{(n)}]dp_c=w[p_c^{(n)}>p_c^{(n-1)}]\cdot g_n[p_c^{(n)}]dp_c,
\ee where $g_n[p_c^{(n)}]$ is the distribution function for $n$-th
percolation threshold, and the probability that the $n$-th threshold
has occurred  after the $(n-1)$-th threshold is equal to
 \be\label{2}
w[p_c^{(n)}>p_c^{(n-1)}]=1-F[p_c^{(n-1)}], \quad
F(t)=\int\limits_0^t f(p_c)dp_c.
 \ee

It follows herefrom
 \be\label{3}
g_n[p_c^{(n)}]= \left\{\begin{tabular}{ll}
$0,\qquad p_c^{(n)}<p_c^{(n-1)}$\\
\\
$\frac{\ds f[p_c^{(n)}]}{\ds 1-F[p_c^{(n-1)}]},\qquad p_c^{(n)}>p_c^{(n-1)}$\\
\end{tabular}\right..
\ee

To calculate the mean value $\bar p_c^{(n)}$ of the $n$-th threshold
one has to perform $n$ averaging procedures   with the help of
functions $g_k[p_c^{(k)}]$ ($k=n, n-1, \ldots, 1$):
$$
\noindent\begin{tabular}{l} $\left<p_c^{(n)}\right>=
\int\limits_{p_c^{(n-1)}}^1 p_c^{(n)}g_{n}[p_c^{(n)}]dp_c^{(n)}=
\frac{\ds \Psi_1[p_c^{(n-1)}]}{\ds 1-F[p_c^{(n-1)}]},
\quad\mbox{where}\quad \Psi_1(t)=\int\limits_t^1 xf(x)dx$,\\
\\
$\left<\left<p_c^{(n)}\right>\right>=
\int\limits_{p_c^{(n-2)}}^1\left<p_c^{(n)}\right>g_{n-1}[p_c^{(n-1)}]dp_c^{(n-1)}=
\frac{\ds\Psi_2[p_c^{(n-2)}]}{\ds
1-F[p_c^{(n-2)}]},\quad\mbox{where}\quad \Psi_2(t)=
\int\limits_t^1\frac{\ds f(x)\Psi_1(x)}{\ds 1-F(x)}dx$,\\
\\
\hbox to \textwidth{\dotfill}\\
\\
$
\underbrace{\left<\ldots\left<p_c^{(n)}\right>\ldots\right>}_{\mbox{\fs
$k$-th averaging}} =\int\limits_{p_c^{(n-k)}}^1
\underbrace{\left<\ldots\left<p_c^{(n)}\right>\ldots\right>}_{\mbox{\fs$(k-1)$-th
averaging}} g_{n+1-k}[p_c^{(n+1-k)}]dp_c^{(n+1-k)}=
\frac{\ds \Psi_{n-k}[p_c^{(n-k)}]}{\ds 1-F[p_c^{(n-k)}]}$,\\
\\
$\hfill\mbox{where}\quad \Psi_{n-k}(t)= \int\limits_t^1\frac{\ds
f(x)\Psi_{n-1-k}(x)}{\ds 1-F(x)}dx,
\qquad$\\
\\
\hbox to \textwidth{\dotfill}\\
\\
$\bar p_c^{(n)}\equiv
\underbrace{\left<\ldots\left<p_c^{(n)}\right>\ldots\right>}_{\mbox{\fs
$n$-th averaging}} =\Psi_n(0), \quad\mbox{where}\quad \Psi_n(0)=
\int\limits_0^1\frac{\ds f(x)\Psi_{n-1}(x)}{\ds 1-F(x)}dx.
\qquad$\\
\end{tabular}
$$
With those recurrent relations one could calculate successively all
mean values $\bar p_c^{(n)}$.

It is difficult to perform analytically this procedure for the
above-found distribution $f(p_c)$, which is close to the Gaussian
one. However, with that distribution taken approximately as the
uniform distribution
 \be\label{4}
f(p_c)=\left\{\begin{tabular}{ll} $\frac{\ds 1}{\ds4\sigma},$ &$|p_c-\bar p_c|<2\sigma$\\
\\
$0$,&$|p_c-\bar p_c|>2\sigma$\\
\end{tabular}\right.
 \ee
of the width $4\sigma$ centered near $p_c=\bar p$, the mentioned
procedure yields the simple result
 \be\label{5}
\bar p_c^{(1)}=\bar p_c,\quad \bar p_c^{(n)}=\bar
p_c+2\sigma\left(1-\frac{1}{2^{n-1}}\right),
 \ee
according to which the mean value of each threshold lies exactly in
the middle between the preceding threshold and the value of
$p_c=\bar p_c+2\sigma$ corresponding to the right boundary of the
adopted uniform distribution.

In the insert of Fig. 3, the values of successive thresholds  $\bar
p_c^{(n)}$ ($n=1,\ldots,6$) are shown for the typical $10\times
1000$-lattice. With increasing $n$-number, thresholds converges
rapidly and their values are well approximated by the Eq. (\ref{5}).
\bigskip

\centerline{\bf 3. Experimental examples}
 \medskip

In this section, we consider some physical phenomena that are
different in all respects excluding the only one: their properties
could be described in the framework of simple percolation models
exploiting the above-derived results.
 \medskip

{\bf Cluster films}. Conducting cluster films are produced by the
generation and the subsequent deposition of small (diameter of
10--100 nm) metal clusters on the insulator substrate with
preliminary deposited contacts. At a certain film thickness
(increasing in the course of the film deposition), the gap between
the contacts becomes conducting. In the experiments~\cite{9,10},
Bi-clusters of the size $60\pm10$~nm have been deposited on the
SiN-substrate. In the course of the deposition, the current $I$
between metal contacts separated by the distance of $h\sim1000$ nm
has been measured, with the contact width of $L\gg h$ and the
voltage drop of $\sim 1$ V.

One of the temporal dependencies $I(t)$ of the current (given
in~\cite{9}) is displayed in Fig. 5. In the course of the
deposition, the fraction $p$ of the surface occupied by the clusters
increases gradually and at one point the current rises steeply (on
three orders of value) that corresponds to originating the first
percolation channel, whereupon the current increases by well-marked
steps of lesser heights corresponding to appearing the next
percolation channels, and finely approaches to saturation. The
insert of Fig. 5 shows the typical result of modeling that process
by calculating the conductance dependence $G(p)$ for one of the
realizations of the $10\times50$-lattice (the site problem for the
lattice with square cells has been solved); to fit the relative
value of the main conductance jump, the value $G_{\rm
min}/G_0=10^{-7}$ has been chosen. The clear resemblance of the
experimental and the model dependencies is indicative of adequacy of
our model for describing the process of the film growth and the
possibility to calculate qualitatively characteristics of that
process\footnote{Some more elaborated model of the film deposition
is considered in~\cite{22}. However, the system of  the square form
where appearing the parallel percolation channels is improbable has
been studied only.}.

Another interesting and important feature of conducting cluster
films (having observed after the deposition completion) is the
current noise indicating of spontaneous temporal alterations of the
film conductance. Fig. 6 shows the typical temporal  dependence
$I(t)$ of the current flowing through the cluster film that
demonstrates the occurrence of the intensive random noise of the
“telegraph” type~\cite{10} against the smooth background of the
conductance change due to the film oxidation (shown by the dash
curve). The relative current variations are of about 10\%. In the
insert of Fig. 6, the typical result of modeling that process is
shown. Conductance variations $G(t^*)$ for one of the realizations
of the $10\times50$-lattice system with $p=0.25$ have been monitored
in the course of commutating status of randomly chosen  sites: the
existent site has been removed, while the absent one has been
recovered. The flow of “time” $t^*$ corresponds to the sequence of
those commutations. Like in the experiment, the relative conductance
variation equals $\sim10\%$. The occurrence of the “telegraph” noise
in the considered percolation system is associated with the
disparity of their sites: some of them are included in the skeleton
cluster determining the average system conductance or close to its
sites, while other sites are apart from them. Therefore, “switching
on” or “switching out” the sites of the first type changes
significantly  the system conductance (in our case, on the order of
$\sim10\%$), while the status of the second type sites influences
the conductance slightly. It is seen from Fig. 6 that the stepwise
variation of the conductance occur  on rare occasions --  only one
attempt among $\sim100$ is “successful”. It means that in the
considered system the current-conducting cluster includes $\sim1\%$
of all sites only.

The insert of Fig. 6 demonstrates also the model temporal dependence
$G(t^*)$ of the conduct\-ance for one of the realizations of the
$10\times50$-lattice system with $p=0.85$. In that case, there is no
telegraph noise, and the relative conductance fluctuations are
significantly less than 10\%. That agree with experimental results
corresponding to the virgin (non-oxidized) cluster films with higher
$p$-value~\cite{10} or -- to observations of the noise in amorphous
$a$-Si-films~\cite{11,12}.
\medskip

{\bf Two-dimensional electron gas in the system with random
electrostatic potential.} The structure
“metal-insulator-semiconductor” is the standard tool for
controllable producing two-dimensional electron gas (near the
interface insulator-semiconductor). Built in such a structure
electric charges (localized in  traps of the gate insulator, for
instance) produce a random electrostatic potential in the region of
two-dimensional electron channel. As this takes place, electrons are
localized in wells of chaotic potential relief and the system turns
into percolation one. The current in such a system flows through
“bonds” that connect potential wells - “sites” and transit mountain
crossings (saddle-points) of the potential relief.

At low electron Fermi energy  $\varepsilon_F$,  those crossings
(relative to the Fermi level) are of high-altitude and, hence, the
conductance is of the activation nature and low. With rising the
Fermi level (being controlled by the gate potential $V_g$), crossing
heights diminish and they gradually “drop out”\footnote{At some
value of the Fermi energy,  the total conductance of the system is
defined by the “weak link” conductance $G_q$ corresponding to the
highest saddle-point on the current path~\cite{8}. Electron flow
through this saddle-point is quasi one-dimensional that (at low
enough temperatures) leads to the quantization of  the conductance
 $G_q= (e^2/\pi\hbar)N\,\,\, (N=1, 2,\ldots
$)~\cite{23}. The present paper does not touch on this interesting
effect.}. In that regime, the conductance is of the percolation
nature and rising the Fermi energy is equivalent to variation of the
percolation parameter $p$ related to the number of “weak links” --
former crossings of the potental relief.

The percolation parameter $p$ could be associated with the relative
fraction of that part of the system surface where $V<\varepsilon_F$,
so
$$
p=\int\limits_{-\infty}^{\varepsilon_F}\phi(V)dV.
$$
Near the percolation threshold $\Delta p\propto\Delta\varepsilon_F$
and, besides, $\partial\varepsilon_F/\partial V_g={\rm
Const}$~\cite{8}. Therefore, the gate voltage $V_g$ and the
percolation parameter $p$ are connected by the linear relationship
$\Delta V_g \propto\Delta p$ permitting the simple comparison of
experimental dependencies $G(V_g)$ with  calculated relations $G(p)$
(see below).

In experiments~\cite{8}, there have been two-dimensional electron
channel with $h=5$ $\mu$m, $L=50$ $\mu$m and the characteristic size
of percolation cluster cells of about 1 $\mu$m that corresponds
approximately to the percolation $10\times 100$-lattice. Fig. 7
compares the experimental dependence $G(V_g)$ with the dependence
$G(p)$ calculated by the above-described method (values
$G_0=2.5\cdot10^{-4}$ Ohm$^{-1}$, and $G_{\rm min}\ll G_0$ have been
accepted). It is seen that both dependencies are well
superimposed\footnote{The existence of steps on the calculated curve
is the consequence of simplifying assumption that conductances of
all bonds are equal. In the real experiment, the spread of those
conductances results in smoothing peculiarities on the
$G(V_g)$-curve.} at $p\gtrsim0.3$ with the preservation of the
linear relationship between $V_g$ and $p$. That argues in favor of
the above-suggested assumption about the percolation nature of the
examined system at high values $V_g$. At lower $V_g$-values the
conductance is the activation one and its description requires
modifying the simple considered model.
\medskip

{\bf Lightning.} The lightning is the mighty short-time electric
discharge in the atmosphere whose length is usually measured by
kilometers.  Lightning discharge is always preceded by the leader
pre-discharge which appears at the  electric field intensity as low
as $\sim 3$ kV/cm that is on the order of magnitude lower than the
electric field necessary for the electric breakdown in air at normal
conditions (30 kV/cm). Various lightning theories overcome that
discrepancy by rather artificial assumptions. Another unusual
feature of the leader is that its  travel to the earth occurs by
steps separated by pauses, so that it has been referred as the
stepped leader. In recent years, there has been suggested the new
hypothesis~\cite{13} of initiating lightnings (producing leaders) by
cosmic rays of ultra-high energies (higher than $10^{12}$~eV).
Charged particles born by the rays (electrons, muons) pass through
the atmosphere and leave behind the system of ionized traces of
different length and various directions -- a kind of a random
\emph{permanently renewed} lattice of conducting bonds, along which
first the leader goes forward and then occurs the lightning
discharge.

Estimates~\cite{13} show that the typical size of the lattice cell
is of about  100~m. With the distance from the bottom surface of the
thunder cloud to the earth of $0.5-3$ km and its length of $10-30$
km, that corresponds to the percolation lattice $h\times
L=10\times(100-500)$

The problem concerning the value of the percolation parameter $p$
involves the special examination. If the system would be stationary,
the rough estimate of $p$ could be obtained on the base of the
following consideration. In addition to single lightning discharges,
simultaneous multiple discharges are frequently observed which could
be associated with multiple percolation channels of above-considered
model. To estimate  the average number of lightning channels
$\langle N\rangle$ for those multiple discharges  we have studied
numerous lightning pictures published in Internet (the total number
of about 350; see, for instance,~\cite{24}) and derived $\langle
N\rangle=3\pm1$. On the other hand, that number is determined by
dimensions of the percolation $h\times L$-lattice and the
percolation parameter $p$. Investigating dependensies similar to
those shown in Fig. 4 (but relating to different realizations of
$10\times300$-lattices) leads to  $\langle N\rangle\approx3$ if one
accepts $p=0.3-0.4$.

However, proceeding from the known intensity of cosmic rays one
could conclude that the obtained estimation is too high: the fluence
of cosmic rays with the needed energy is sufficient for producing
(in the moment) the system of ionized traces with $p\sim0.01$.
According to Fig. 1, the probability of generating  conducting
channel in that system is negligible, i.e., the lighning would be
absolutely unique phenomenon. Nature overcomes that difficulty
executing the breakdown of the air gap between the cloud and earth
by means of the  \emph{stepped} leader discharge. Each successive
step of its expansion occurs after the pause during which the leader
waits for generating (by cosmic rays) a new ionized trace allowing
to go forward. In the context of the percolation theory, there
arises the new problem concerning the delayed, or stepped,
percolation in the \emph{time-dependent} lattice of bonds.
\bigskip

\centerline{\bf 4. Conclusions}
 \medskip

March to new technologies allowing the successive diminishing
characteristic dimensions of electronic devices transforms the
latter into mesoscopic systems whose properties are identical on the
average only. Many of them are intrinsically finite-scale
percolation systems whose characteristic dimensions exceed the size
of constituent elements by 10--100 times (and on occasion -- by
several times only). In the present paper, some features of such
two-dimensional  systems have been studied. In particular, there
have been  investigated statistical properties of their percolation
thresholds, defined successive thresholds of arising multiple
percolation channels, calculated the conductance dependence of such
systems on their dimensions and percolation parameter $p$. Special
attention has been given to systems whose dimension along the
direction of the current flow is significantly smaller than in the
direction perpendicular to the current. It has been shown that
percolation finite-scale systems are natural mathematical models
suitable for  the qualitative and quantitative description of
different physical systems.

This work was supported by Grants 03-02-17029, 05-02-17021 of the
Russian Foundation of Basic Researches.
\newpage

\centerline{\bf Figure captions} \vspace{1cm}

Fig. 1. Distribution function $f(p_c)$ of percolation thresholds
(site problem) for random rectangular lattices with square cells
defined with numerical calculations of thresholds for $10^5$
realizations of each system (threshold for the infinite lattice is
$p_{c\infty}\approx0.59$). Marking of distributions corresponds to
lattice dimensions in the format  $h\times L$. Curves are
approximations of distribution functions by Gaussian ones. \\

Fig. 2. Parameters of Gaussian functions approximating the
distribution functions $f(p_c)$ of percolation thresholds for random
lattices of different dimensions shown in Fig. 1. Three top curves
-- mean threshold values $\bar p_c$, three bottom curves -- widths
of  distributions $\sigma$.\\

Fig. 3. Typical dependence $G(p)$ of the transversal conductance of
a random $10\times1000$-lattice in the form of the long strip on the
$p$-value. Arrows indicate specific $p$-values (successive
percolation thresholds  $p_c^{(n)}$) corresponding to originating
new conducting channels. In the insert -- $p_c^{(n)}$-values
corresponding to the dependence $G(p)$ shown ({\Large$\bullet$}) and
the analytical relation (\ref{5}) ({\Large$\circ$}).\\

Fig. 4. Dependence $G(L)$ of the transversal conductance of a
typical random lattice of increasing length with $h=10$ and
$1<L\le1000$ on its length $L$ at $p=\mbox{Const}$. Each jump
corresponds to linking up the section containing the conducting
channel. \\

Fig. 5.  Temporal  dependence of the current $I(t)$ through the
cluster film in the course of its deposition~\cite{9}. In the insert
-- dependence of the $10\times50$-lattice conductance $G(p)$ with
the ratio $G_{\rm min}/G_0=10^{-7}$.\\

Fig. 6. Temporal dependence of the current $I(t)$ through the
cluster film after terminating the process of its deposition (the
voltage equals 5 mV, $T=300$~K)~\cite{10}. In the insert --
“temporal” dependences $G(t^*)$ of  $10\times50$-lattice conductance
for  $p=0.25$ and $p=0.85$.\\

Fig. 7. Experimental conductance dependence $G(V_g)$ for
two-dimensional electron gas in the electrostatically disordered
structure “metal-insulator-semiconductor” ($T=300$ K) on the gate
voltage~\cite{8} (dotted curve; bottom and left coordination axes)
and the typical calculated conductance dependence $G(p)$ for the
percolation  $10\times100$-lattice (bond problem) on the percolation
parameter $p$ (dots; top and right coordination axes).

\newpage

$^*$Electronic address: meilikhov@imp.kiae.ru 

\newpage
\begin{figure}
\includegraphics[width=0.9\textwidth]{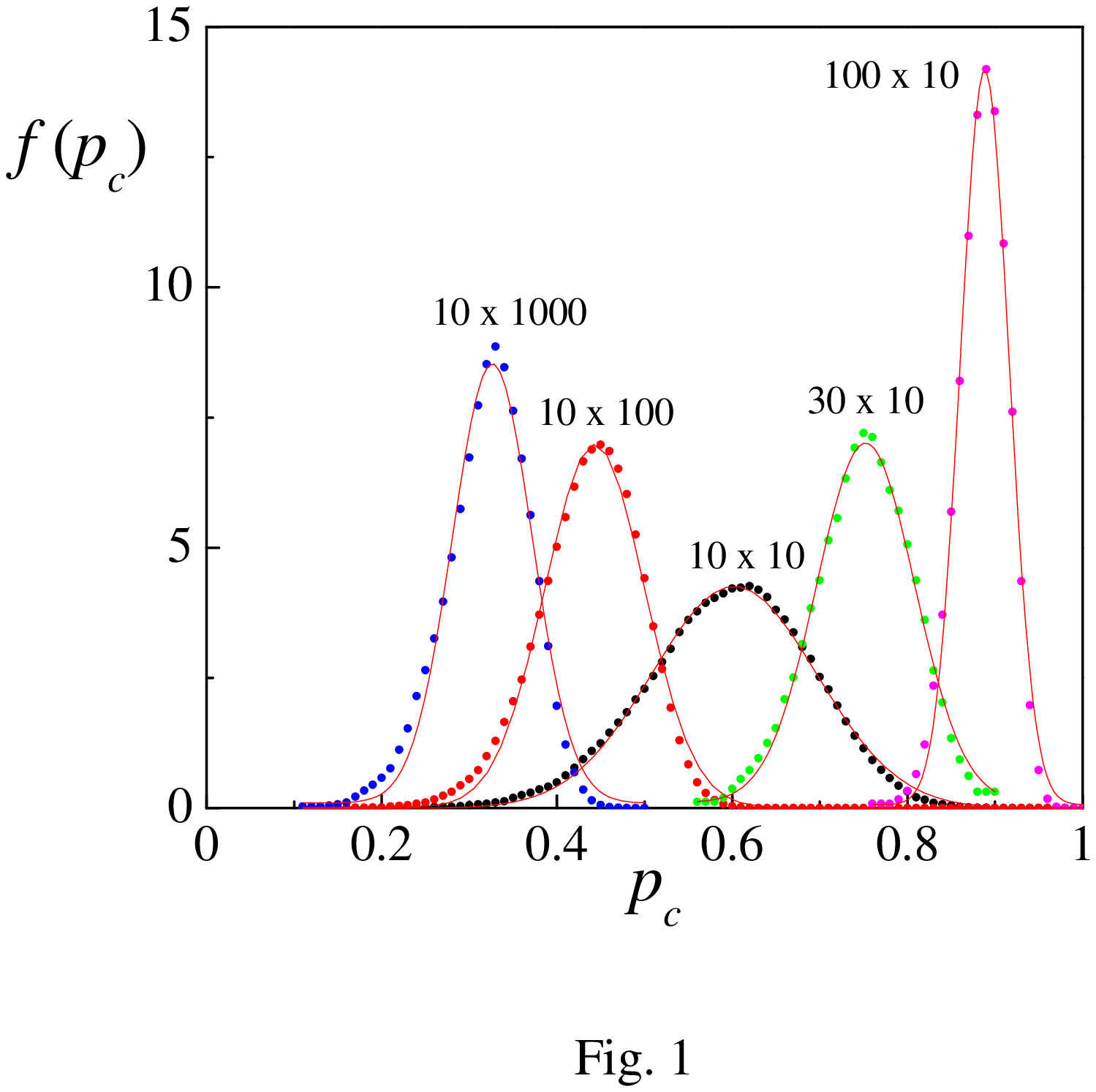}
\end{figure}

\newpage
\begin{figure}
\includegraphics[width=0.9\textwidth]{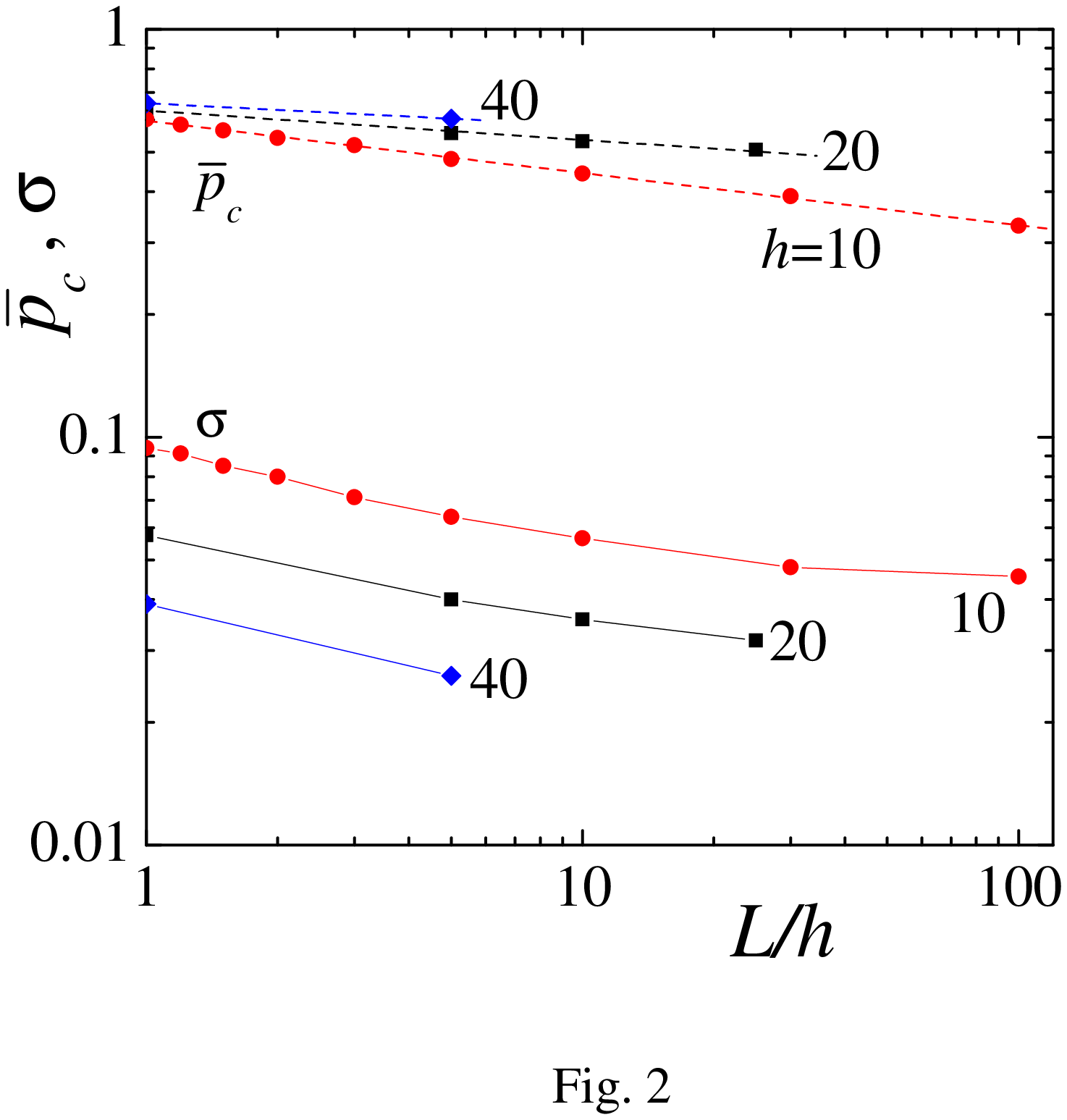}
\end{figure}

\newpage
\begin{figure}
\includegraphics[width=0.9\textwidth]{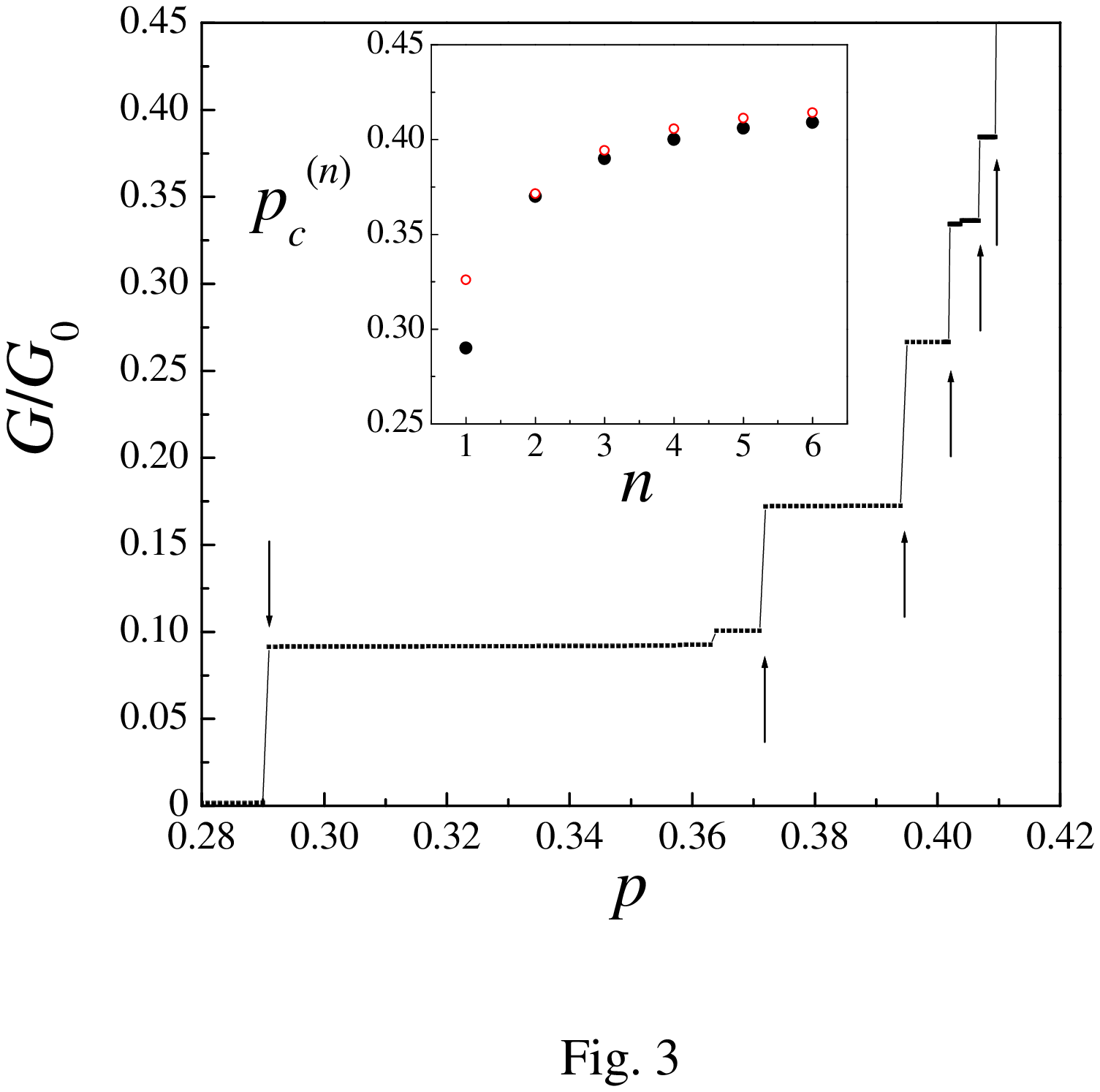}
\end{figure}

\newpage
\begin{figure}
\includegraphics[width=0.9\textwidth]{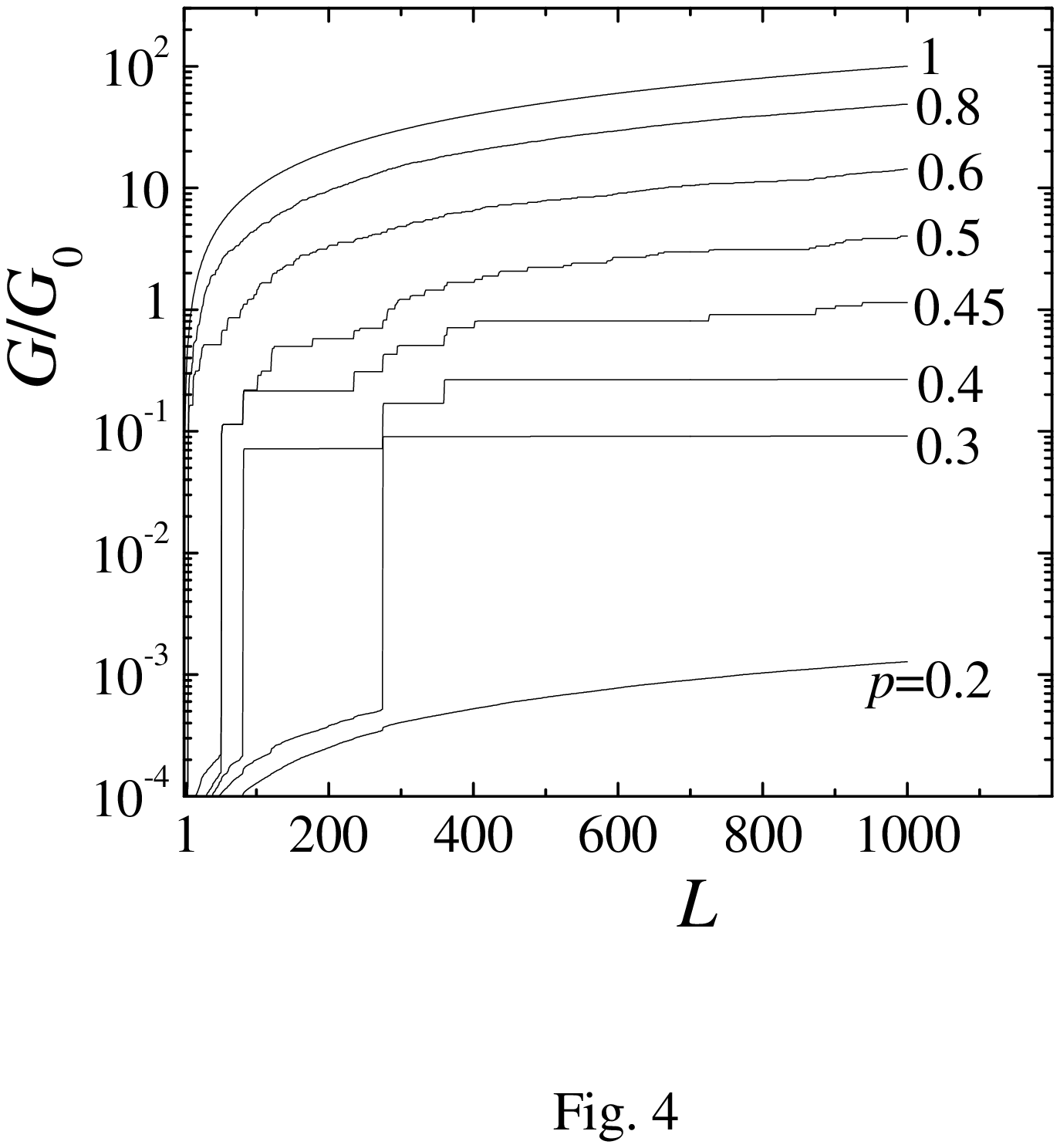}
\end{figure}

\newpage
\begin{figure}
\includegraphics[width=0.9\textwidth]{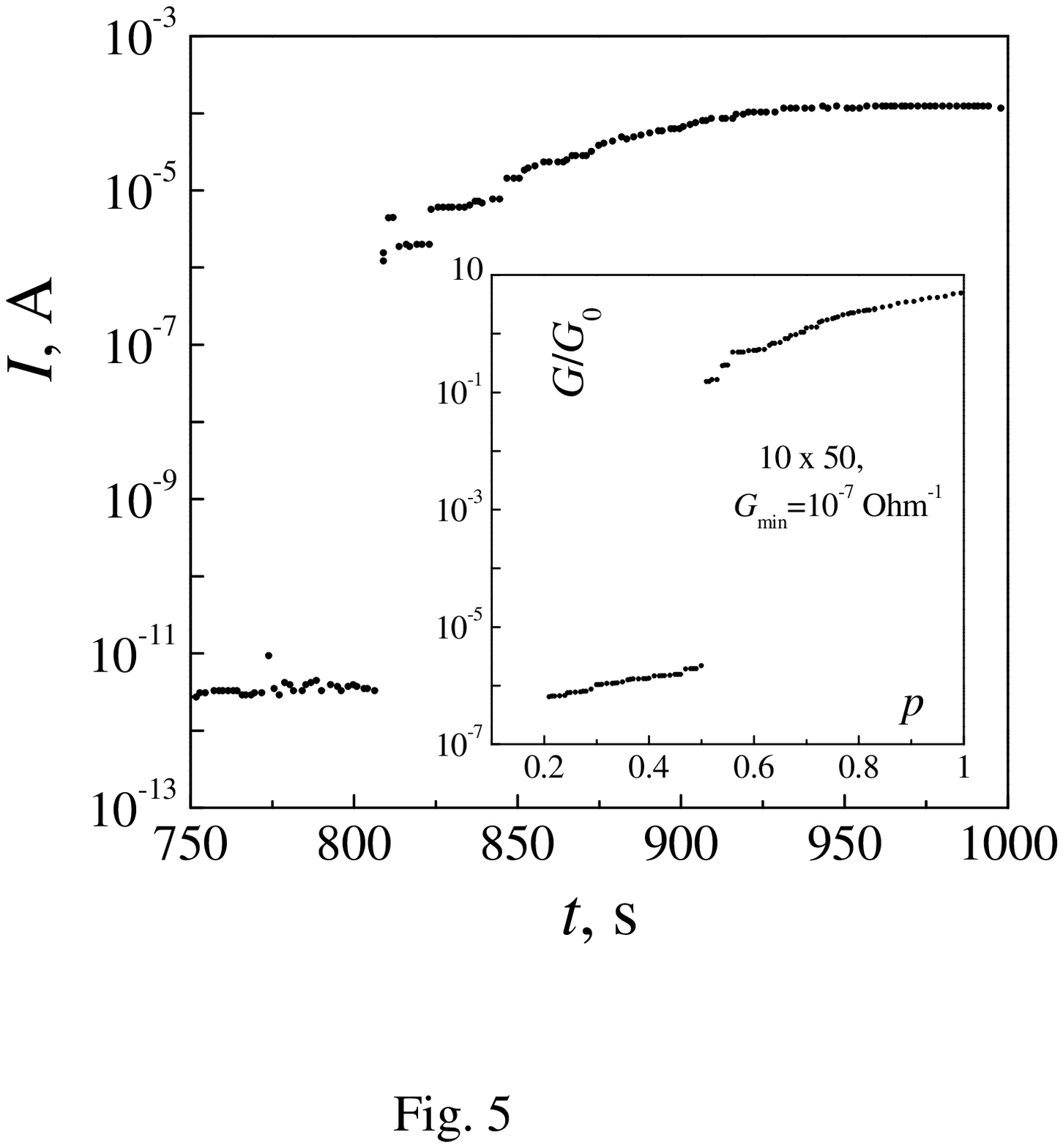}
\end{figure}

\newpage
\begin{figure}
\includegraphics[width=0.9\textwidth]{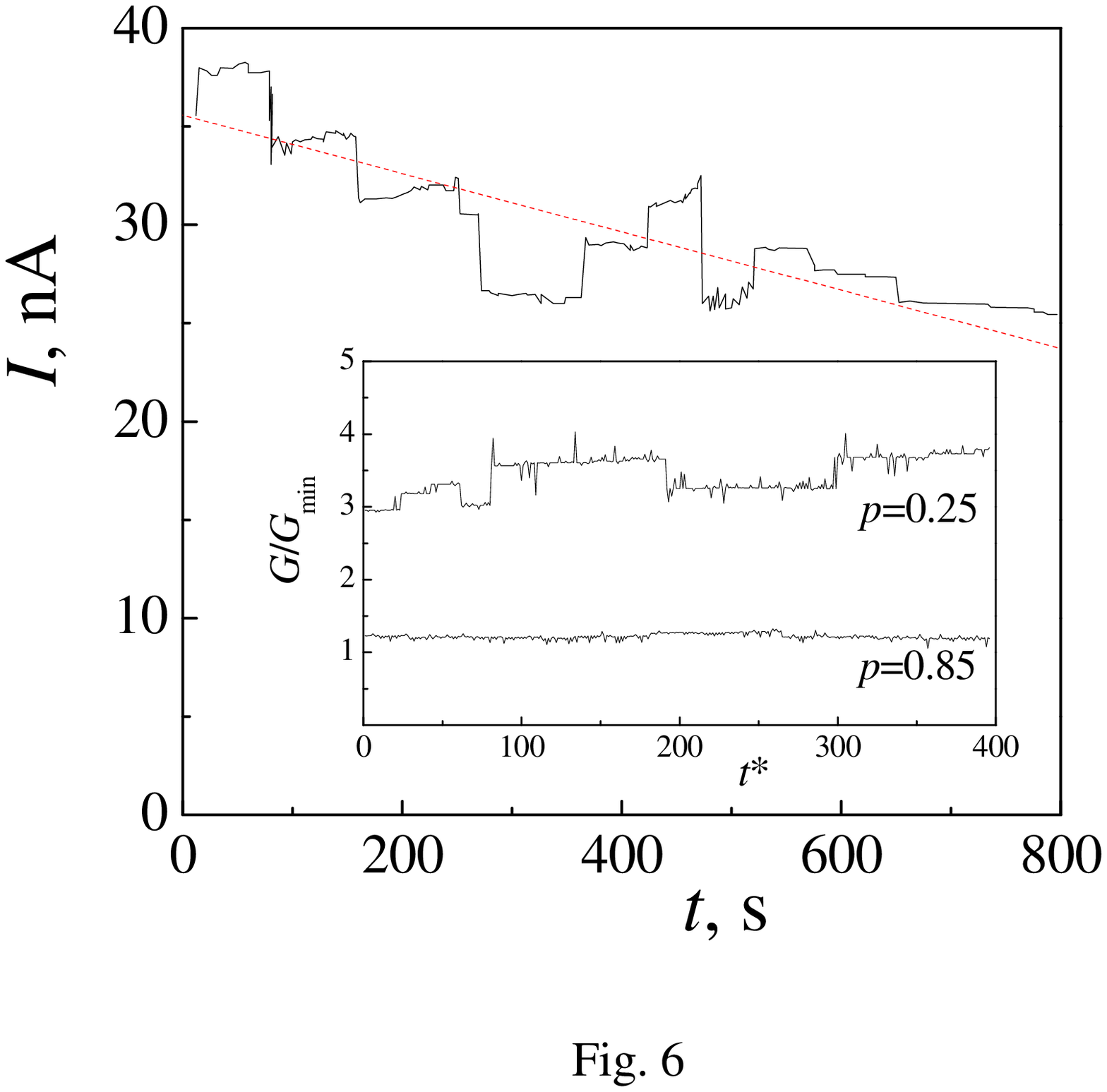}
\end{figure}

\newpage
\begin{figure}
\includegraphics[width=0.9\textwidth]{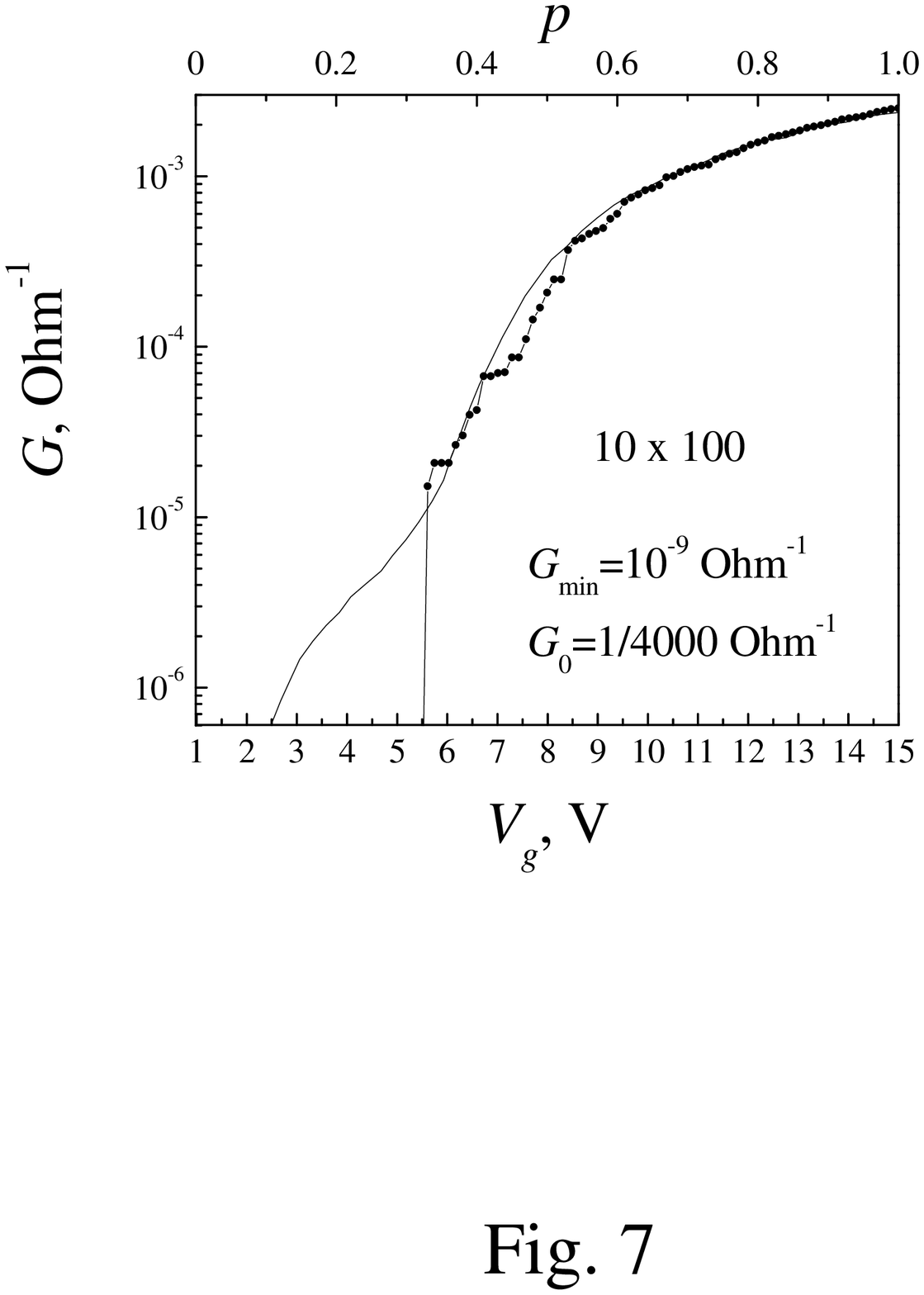}
\end{figure}


\vspace{-42pt}
\vspace{1cm}

\begin{thebibliography}{30}
\bibitem{1} D. Stauffer, A. Aharony, \emph{Introduction to Percolation
Theory} (Tailor\& Francis, London, 1994).
\bibitem{2} R. Langlands, C. Pichet, P. Pouliot, and Y. Saint-Aubin,
Stat. Phys., {\bf 67}, 553 (1992).
\bibitem{3} M. Aisenman, Nucl. Phys. B, {\bf 485}, 551 (1997).
\bibitem{4} C.-K. Hu, and C.-Y. Lin, Phys. Rev. Lett., {\bf 77}, 8
(1996).
\bibitem{5} J. Cardy, J. phys. A, {\bf 31}, L105 (1998)
\bibitem{6} R. P. Langlands, C. Pichet, Ph. Pouliot, Y. Saint-Aubin,
J. Stat. Phys., {\bf 67}, 553 (1992).
\bibitem{7}  J. H. Cardy, J. Phys. A, {\bf 25}, L201 (1992).
\bibitem{8}  B. A. Aronzon, D. A. Bakaushin, A. S. Vedeneev, A. B. Davydov, E. Z. Meilikhov,
 and N.~K. Chumakov,  Semiconductors, {\bf 35}, 436 (2001).
\bibitem{9} J. Schmelzer, Jr., S. A. Brown, A. Wurl, M. Hyslop, R. J. Blaikie,
Phys. Rev. Lett., {\bf 88}, 226802 (2002).
\bibitem{10} M. Schulze, S. Gourley, S.A. Brown, A. Dunbar, J.
Partridge, R. J. Blaikie, Eur.~Phys.~J.~D, {\bf 24}, 291 (2003).
\bibitem{11} C. E. Parman, N. E. Israeloff, J. Kakalios, Phys. Rev.~B, {\bf 47}, 12578
(1993).
\bibitem{12} L. M. Lust, J. Kakalios, Phys. Rev. Lett., {\bf 75}, 2192
(1995).
\bibitem{13} V.I. Ermakov, Y.I. Stozhkov, Proc. of 11th Int. Conf. on Atmospheric Electricity,
Alabama, USA, 242 (1999).
\bibitem{14} D. Stauffer, Physica A, {\bf 242}, 1 (1997).
\bibitem{15} B.I. Shklovskii, A.L. Efros, \emph{Electronic properties of doped semiconductors}
(Springer, Heidelberg, 1984).
\bibitem{16} C.-K. Hu, Phys. Rev. B, {\bf 51}, 3922 (1995).
\bibitem{17} C.-K. Hu, C.-Y. Lin, J.-A. Chen, Phys. Rev. Lett., {\bf 75}, 193
(1995).
\bibitem{18} S. Tsubakihara, Phys. Rev. E, {\bf 62}, 8811 (2000).
\bibitem{19} R.A. Monetti, E.V. Albano, Z. Phys. B, {\bf 90}, 351
(1993).
\bibitem{20} C.-K. Hu, C.-Y. Lin, J.A. Chen, Phys. Rev. Lett., {\bf 75}, 193
(1995).
\bibitem{21} B. Derrida, J. Vannimenus, J. Phys. A, {\bf 15}, L557
(1982).
\bibitem{22} N.I. Lebovka, S.S. Manna, S. Tarafdar, and N.
Teslenko, Phys. Rev. E, {\bf 66}, 066134 (2202).
\bibitem{23} M. Buttiker, Phys. Rev. B, {\bf 41}, R7906 (1990).
\bibitem{24} http://www.strikingimages.com/light.htm

\end{thebibliography}
\end{document}